\input amstex
\magnification 1200
\TagsOnRight
\def\qed{\ifhmode\unskip\nobreak\fi\ifmmode\ifinner\else
 \hskip5pt\fi\fi\hbox{\hskip5pt\vrule width4pt
 height6pt depth1.5pt\hskip1pt}}
 \def\adots{\mathinner{\mkern2mu\raise1pt\hbox{.}
\mkern3mu\raise4pt\hbox{.}\mkern1mu\raise7pt\hbox{.}}}
\def\sdots{\mathinner{
     \mskip.01mu\raise1pt\vbox{\kern1pt\hbox{.}}
     \mskip.01mu\raise3pt\hbox{.}
     \mskip.01mu\raise5pt\hbox{.}
     \mskip1mu}}
\NoBlackBoxes
\baselineskip 20 pt
\parskip 5 pt

\centerline {\bf EXACT SOLUTIONS TO THE}
\centerline {\bf NONLINEAR SCHR\"ODINGER EQUATION}

\vskip 10 pt
\centerline {Tuncay Aktosun and Theresa Busse}
\vskip -6 pt
\centerline {Department of Mathematics}
\vskip -6 pt
\centerline {University of Texas at Arlington}
\vskip -6 pt
\centerline {Arlington, TX 76019-0408, USA}

\centerline{Francesco Demontis and Cornelis van der Mee}
\vskip -6 pt
\centerline{Dipartimento di Matematica e Informatica}
\vskip -6 pt
\centerline{Universit\`a di Cagliari}
\vskip -6 pt
\centerline{Viale Merello 92}
\vskip -6 pt
\centerline{09123 Cagliari, Italy}

\vskip 10 pt

\noindent {\bf Abstract}:
A review of a recent method is presented to construct certain exact solutions to the
focusing nonlinear Schr\"odinger equation on the line
with a cubic nonlinearity. With motivation by the inverse scattering transform
and help from the state-space method,
an explicit
formula is obtained to express such
exact solutions in a compact form in terms of a
matrix triplet and by using matrix
exponentials. Such solutions consist of
multisolitons with any multiplicities, are analytic on the entire
$xt$-plane, decay exponentially as $x\to\pm\infty$ at each fixed $t,$ and can alternatively
be written explicitly as algebraic combinations of
exponential, trigonometric, and polynomial functions
of the spatial and temporal coordinates $x$ and $t.$
Various equivalent forms of the matrix triplet are presented
yielding the same exact solution.


\newpage

\noindent {\bf 1. INTRODUCTION}
\vskip 3 pt

Our goal in this paper is to review and further elaborate on
a recent method [3,4] to construct certain exact
solutions
to the focusing nonlinear Schr\"odinger (NLS) equation
$$iu_t+u_{xx}+2|u|^2u=0,\tag 1$$
with a cubic nonlinearity, where the subscripts denote
the corresponding partial derivatives.

The
NLS equation has important applications in various areas
such as wave propagation in nonlinear
media [15], surface waves on deep waters [14], and
signal propagation in optical fibers [9-11]. It was the second
nonlinear partial differential equation (PDE) whose initial value
problem was discovered [15] to be solvable via the inverse
scattering transform (IST) method.
Recall that the IST method associates (1)
with the Zakharov-Shabat system
$$\displaystyle\frac{d \varphi(\lambda,x,t)}{dx}=\bmatrix -i\lambda&u(x,t)\\
\noalign{\medskip} -u(x,t)^\ast&i\lambda\endbmatrix
\varphi(\lambda,x,t),\tag 2$$ where $u(x,t)$ appears as a
potential and an asterisk is used for complex conjugation. By
exploiting the one-to-one correspondence between the potential
$u(x,t)$ and the corresponding scattering data for (2), that
method amounts to determining the time evolution $u(x,0)\mapsto
u(x,t)$ in (1) with the help of solutions to the direct and
inverse scattering problems for (2).

Even though we are motivated by the IST method, our goal is not
to solve the initial value problem for (1). Our aim is rather
to construct certain exact solutions to (1) with the help of a
matrix triplet and by using matrix exponentials. Such exact
solutions turn out to be multisolitons with any multiplicities.
Dealing with even a single soliton with multiplicities has not
been an easy task in other methods; for example, the exact
solution example presented in [15] for a one-soliton solution
with a double pole, which is obtained by coalescing two
distinct poles into one, contains a typographical error, as
pointed out in [13].

In constructing our
solutions we make use of the state-space method [6] from the control theory.
Our solutions are uniquely constructed via the explicit formula (9),
which uses as input
three (complex) constant matrices $A,$ $B,$ $C,$ where
$A$ has size $p\times p,$ $B$ has size $p\times 1,$ and
$C$ has size $1\times p,$ with $p$ as any positive integer.
We will refer to $(A,B,C)$ a triplet of size $p.$ There is no loss of generality
in using a triplet yielding a minimal representation [3,4,6], and we will only
consider such triplets.
As seen from the explicit formula (9), our solutions are well defined
as long as the matrix $F(x,t)$ defined in (8) is invertible. It turns out that
$F(x,t)$ is invertible if and only if two conditions are met on
the eigenvalues of the constant matrix $A;$
namely, none of the eigenvalues of $A$ are purely imaginary and
that no two eigenvalues of $A$ are symmetrically located
with respect to the imaginary axis.
Our solutions given by (9) are globally analytic on the entire
$xt$-plane and decay exponentially as $x\to\pm\infty$
for each fixed $t\in\bold R$ as long as those two conditions
on the eigenvalues of $A$ are satisfied.

In our method [3,4] we are motivated by using the IST with rational
scattering data. For this purpose we exploit
the state-space method [6];
namely, we use a matrix triplet $(A,B,C)$ of an appropriate size
in order to represent a rational function vanishing at infinity
in the complex plane.
Recall that any rational function $R(\lambda)$ in the complex plane that vanishes
at infinity has a matrix realization in terms of a matrix
triplet $(A,B,C)$ as
$$R(\lambda)=-iC(\lambda I-i A)B,\tag 3$$
where $I$ denotes the identity matrix.
The smallest integer $p$ in the size of the triplet
yields a minimal realization for $R(\lambda)$ in (3). A
minimal realization is unique up to a similarity
transformation. The poles of $R(\lambda)$ coincide with the eigenvalues
of $(iA).$

The use of a matrix realization in the IST method allows us to establish
the separability of the kernel of a related Marchenko integral
equation [1,2,4,12] by expressing that kernel in terms of a matrix exponential. We then
solve that Marchenko integral equation algebraically and observe
that our procedure leads to exact solutions to the NLS equation even
when the input to the Marchenko equation does not necessarily come
from any scattering data. We refer the reader to [3,4] for details.

The explicit formula (9) provides a compact and concise way to express our exact solutions.
If such solutions are desired to be expressed
in terms of exponential, trigonometric (sine and cosine), and polynomial functions of
$x$ and $t,$ this can also be done explicitly and easily
by ``unpacking" matrix exponentials in (9).
If the size $p$ in the matrices $A,$ $B,$ $C$ is larger than $3,$
such expressions become long; however,
we can still explicitly evaluate them for any matrix size $p$ either by hand or by using
a symbolic software package such as Mathematica.
The power of our method is that we can produce exact solutions
via (9) for any positive integer $p.$
In some other available methods, exact solutions are usually tried to be produced directly
in terms of elementary functions
without using matrix exponentials, and hence any concrete examples that can be produced by
such other methods will be relatively simple
and we cannot expect those other methods to produce our exact solutions
when $p$ is large.

Our method is generalizable to obtain similar
explicit formulas for exact solutions to other integrable nonlinear
PDEs where the IST involves the use of a Marchenko integral
equation [1,2,4,12]. For example, a similar method has been used [5] for the
half-line Korteweg-de Vries equation, and it can be applied to other
equations such as
the modified Korteweg-de Vries equation and the sine-Gordon
equation.
Our method is also generalizable to the matrix versions of
such integrable nonlinear PDEs. For instance, a similar
method has been applied in the third author's Ph.D. thesis [8] to
the matrix NLS equation in the focusing case with a cubic nonlinearity.

Our method also easily handles nonsimple bound-state poles
and the time evolution of the corresponding bound-state norming
constants. In the literature, nonsimple bound-state poles are
usually avoided due to mathematical complications. We refer the
reader to [13], where nonsimple bound-state poles were investigated
and complications were encountered. A systematic treatment of
nonsimple bound states has recently been given in the second author's Ph.D. thesis [7].

Our main results are summarized in Section 2 and some explicit
examples are provided in Section 3. For the proofs, further
results, details, and a summary of other methods to solve the
NLS equation exactly, we refer the reader to [3,4].

\vskip 10 pt
\noindent {\bf 2. MAIN RESULTS}
\vskip 3 pt

In this section we summarize our method to construct certain
exact solutions to the NLS equation in terms of a given triplet $(A,B,C)$
of size $p.$
For the
details of our method we refer the reader to [3,4]. Without any loss of
generality, we assume
that our starting triplet $(A,B,C)$ corresponds to a minimal realization in (3).
Let us use
a dagger to denote the matrix adjoint (complex conjugate and
matrix transpose), and let
the set $\{a_j\}_{j=1}^m$ consist of
the distinct eigenvalues of $A,$ where the algebraic
multiplicity of each eigenvalue may be greater than one and we use $n_j$ to denote
that multiplicity.  We only impose the restrictions
that no $a_j$ is purely imaginary and that no two distinct $a_j$ values are located
symmetrically with respect to the imaginary axis
on the complex plane. Let us set $\lambda_j:=ia_j$ so that we can equivalently
state our restrictions as that no $\lambda_j$ will be real and no two distinct
$\lambda_j$ values will be complex conjugates of each other.
Our method uses the
following steps:

\item{(i)} First construct the constant $p\times p$ matrices
$Q$ and $N$ that are the unique solutions, respectively, to
the Lyapunov equations
$$Q\,A+A^\dagger Q=C^\dagger C,\tag 4$$
$$A\,N+N\,A^\dagger=BB^\dagger.\tag 5$$
In fact, $Q$ and $N$ can be written explicitly in terms of
the triplet $(A,B,C)$ as
$$Q=\displaystyle\frac{1}{2\pi}\int_{\gamma}
d\lambda\,(\lambda I+iA^\dagger)^{-1}C^\dagger C(\lambda I-iA)^{-1},\tag 6$$
$$N=\displaystyle\frac{1}{2\pi}\int_{\gamma}
d\lambda\,(\lambda I-iA)^{-1}BB^\dagger(\lambda I+iA^\dagger)^{-1},\tag 7$$
where $\gamma$ is any positively oriented simple closed contour enclosing
all $\lambda_j$ in such a way that all $\lambda_j^*$
lie outside $\gamma.$ The existence and uniqueness of
the solutions to (4) and (5) are assured by the fact that $\lambda_j\ne \lambda_j^\ast$
for all $j=1,2,\dots,m$ and $\lambda_j\ne \lambda_k^\ast$ for $k\ne j.$

\item{(ii)} Construct the $p\times p$ matrix valued function $F(x,t)$ as
$$F(x,t):=e^{2A^\dagger x-4i(A^\dagger)^2t}+Q
\,e^{-2Ax-4iA^2t}N.\tag 8$$

\item{(iii)} Construct the scalar function $u(x,t)$ via
$$u(x,t):=-2B^\dagger F(x,t)^{-1}
C^\dagger.\tag 9$$
Note that $u(x,t)$ is uniquely constructed from the
triplet $(A,B,C).$ As seen from (9), the quantity
$u(x,t)$ exists at any point on the $xt$-plane
as long as the matrix $F(x,t)$ is invertible.
It turns out that
$F(x,t)$ is invertible on the entire $xt$-plane
as long as
$\lambda_j\ne \lambda_j^\ast$
for all $j=1,2,\dots,m$
and $\lambda_j\ne \lambda_k^\ast$ for $k\ne j.$

Let us note that the matrices $Q$ and $N$ given in (6) and (7)
are known in control theory as the observability Gramian and
the controllability Gramian, respectively, and that it is well
known in control theory that (6) and (7) satisfy (4) and (5),
respectively. In the context of system theory, the
invertibility of $Q$ and $N$ is described as the observability
and the controllability, respectively. In our case, both $Q$ and $N$ are invertible
due to
the appropriate restrictions imposed on the triplet
$(A,B,C),$ which we will
see in Theorem 1 below.

Our main results are summarized in the following theorems. For
the proofs we refer the reader to [3,4]. Although the results
presented in Theorem 1 follow from the results in the
subsequent theorems, we state Theorem 1 independently to
clearly illustrate the validity of our exact solutions to the
NLS equation.

\noindent {\bf Theorem 1.} {\it Consider any triplet $(A,B,C)$ of size $p,$
corresponding to a minimal representation in (3),
and assume that
none of the eigenvalues of $A$ are purely imaginary and
that no two eigenvalues of $A$ are symmetrically located
with respect to the imaginary axis. Then:}

\item{(i)} {\it The Lyapunov equations (4) and (5) are uniquely solvable, and
their solutions are given by (6) and (7), respectively.}
\item{(ii)} {\it The constant matrices $Q$ and $N$
given in (6) and (7), respectively, are selfadjoint; i.e.
$Q^\dagger=Q$ and $N^\dagger=N.$ Furthermore, both $Q$ and $N$ are invertible.}

\item{(iii)} {\it The matrix $F(x,t)$ defined in (8) is
    invertible on the entire $xt$-plane, and the function
    $u(x,t)$ defined in (9) is a solution to the NLS equation
    everywhere on the $xt$-plane. Moreover, $u(x,t)$ is
    analytic on the entire $xt$-plane and it decays
    exponentially as $x\to\pm\infty$ at each fixed $t\in\bold
    R.$}

We will say that two triplets $(A,B,C)$ and
$(\tilde A,\tilde B,\tilde C)$ are equivalent if they yield the same
potential $u(x,t)$ through (9).
The following result shows that, as far as constructing solutions via (9) is concerned,
there is no loss of generality is choosing
our starting triplet $(A,B,C)$ of size $p$ so that it
corresponds to a minimal representation in (3) and
that
all eigenvalues $a_j$
of the matrix $A$ have positive real parts.

\noindent {\bf Theorem 2.} {\it
Consider any triplet $(\tilde A,\tilde B,\tilde C)$ of size $p,$
corresponding to a minimal representation in (3),
and assume that
none of the eigenvalues of $\tilde A$ are purely imaginary and
that no two eigenvalues of $\tilde A$ are symmetrically located
with respect to the imaginary axis. Then,
there exists an equivalent
triplet $(A,B,C)$ of size $p,$
corresponding to a minimal representation in (3) in such a way that
all eigenvalues of $A$ have positive real parts.}

The next two result given in Theorems 3 and 4 show some of the
advantages of using a triplet $(A,B,C)$ where all eigenvalues
of $A$ have positive real parts. Concerning Theorem 2, we
remark that the triplet $(A,B,C)$ can be obtained from $(\tilde
A,\tilde B,\tilde C)$ and vice versa with the help of Theorem 5
or Theorem 6 given below.

\noindent {\bf Theorem 3.} {\it Consider any triplet
$(A,B,C)$ of size $p,$
corresponding to a minimal representation in (3).
Assume that
all eigenvalues of $A$ have positive real parts. Then:}

\item{(i)} {\it The solutions $Q$ and $N$ to (4) and (5),
respectively, can be expressed in terms of the triplet
$(A,B,C)$ as}
$$Q=\int_0^\infty ds\,[Ce^{-As}]^\dagger [Ce^{-As}],\qquad
N=\int_0^\infty ds\,[e^{-As}B][e^{-As}B]^\dagger.\tag 10$$
\item{(ii)} {\it $Q$ and $N$ are invertible, selfadjoint
    matrices.}
\item{(iii)} {\it Any square submatrix of $Q$ containing
the (1,1)-entry or $(p,p)$-entry of $Q$ is invertible. Similarly,
any square submatrix of $N$ containing
the (1,1)-entry or $(p,p)$-entry of $N$ is invertible.}

\noindent {\bf Theorem 4.} {\it Consider a triplet $(\tilde A,\tilde B,\tilde C)$ of size $p,$
corresponding to a minimal representation in (3) and
that
all eigenvalues $a_j$
of the matrix $\tilde A$ have positive real parts
and that the multiplicity of
$a_j$ is
$n_j.$ Then, there exists an equivalent
triplet $(A,B,C)$ of size $p$
corresponding to a minimal representation in (3) in such a way that
$A$ is in a Jordan canonical form with each Jordan block
containing a distinct eigenvalue $a_j$ and having $-1$ in the superdiagonal entries, and
the entries of $B$ consist of zeros and ones. More specifically, we have}
$$A=\bmatrix A_1&
0&\dots&0\\
0& A_2&\dots&0\\
\vdots&\vdots &\ddots&\vdots\\
0&0&\dots&A_m\endbmatrix,
\quad
B=\bmatrix
B_1\\
B_2\\
\vdots\\
B_m\endbmatrix,\quad
C=\bmatrix C_1&C_2& \dots
&
C_m\endbmatrix,\tag 11$$
$$A_j:=\bmatrix a_j&
-1&0&\dots&0\\
0& a_j& -1&\dots&0\\
0&0&a_j& \dots&0\\
\vdots&\vdots &\vdots &\ddots&\vdots\\
0&0&0&\dots&a_j\endbmatrix,
\quad
B_j:=\bmatrix
0\\
0\\
\vdots\\
0\\
1\endbmatrix, \quad C_j:=\bmatrix c_{j(n_j-1)}& \dots & c_{j
1}& c_{j 0}\endbmatrix,$$ {\it where $A_j$ has size $n_j\times
n_j,$ $B_j$ has size $n_j\times 1,$ $C_j$ has size $1\times
n_j,$ and the (complex) constant $c_{j(n_j-1)}$ is nonzero.}

We will refer to the specific form of the triplet
$(A,B,C)$ given in (11) as a standard form.

The transformation between two equivalent triplets can be obtained
with the help of the following two theorems.
First, in Theorem 5 below we consider the transformation where
all
eigenvalues of $A$ are reflected with respect to the imaginary axis.
Then, in Theorem 6 we consider transformations
where only some of the eigenvalues
of $A$ are reflected with respect to the imaginary axis.

\noindent{\bf Theorem 5.} {\it Assume that the triplet $(A,B,C)$ of size $p$
corresponds to a minimal realization in (3) and that all eigenvalues of $A$
have positive real parts. Consider the transformation}
$$(A,B,C,Q,N,F)\mapsto(\tilde A,\tilde B,\tilde C,
\tilde Q,\tilde N,\tilde F),\tag 12$$
{\it where $(Q,N)$ corresponds to the unique
solution to the Lyapunov system in (4) and (5),
the quantity $F$ is as
in (8),}
$$\tilde A=-A^\dagger,\quad \tilde B=-N^{-1}B,\quad
\tilde C=-CQ^{-1},\quad \tilde Q=-Q^{-1},\quad \tilde N=-N^{-1},$$
{\it and $\tilde F$ and $\tilde u$
are as in (8) and (9), respectively,
but by using $(\tilde A,\tilde B,\tilde C,\tilde Q,\tilde N)$ instead of $(A,B,C,Q,N)$
on the right hand sides. We then have the following:}
\item{(i)} {\it The matrices $\tilde Q$ and $\tilde N$ are selfadjoint and invertible.
They satisfy the respective
Lyapunov equations}
$$\cases \tilde Q\tilde A+\tilde A^\dagger \tilde Q=\tilde C^\dagger \tilde C,\\
\noalign{\medskip}
\tilde A\tilde N+\tilde N\tilde A^\dagger=\tilde B\tilde B^\dagger.\endcases\tag 13$$

\item{(ii)} {\it The quantity $F$ is transformed as
$\tilde F=Q^{-1}FN^{-1}.$ The matrix $\tilde F$ is invertible
at every point on the $xt$-plane.}

To consider the case where only some of eigenvalues of $A$ are
reflected with respect to the imaginary axis, let us again
start with a triplet $(A,B,C)$ of size $p$ and corresponding to
a minimal realization in (3), where the eigenvalues of $A$ all
have positive real parts. Without loss of any generality, let
us assume that we partition the matrices $A,$ $B,$ $C$ as
$$A=\bmatrix A_1&
0\\
\noalign{\medskip}
0&A_2\endbmatrix,\quad B=\bmatrix B_1\\
\noalign{\medskip}
B_2\endbmatrix,\quad C=\bmatrix C_1&C_2\endbmatrix,\tag 14$$
so that the $q\times q$ block diagonal matrix $A_1$ contains the eigenvalues
that will remain unchanged and
$A_2$ contains the eigenvalues
that will be reflected with respect to the imaginary axis
on the complex plane, the submatrices $B_1$ and $C_1$ have sizes
$q\times 1$ and $1\times q,$ respectively, and hence
$A_2,$ $B_2,$ $C_2$ have sizes
$(p-q)\times(p-q),$ $(p-q)\times 1,$ $1\times (p-q),$ respectively,
for some integer $q$ not exceeding $p.$
Let us clarify our notational choice in (14) and emphasize that
the partitioning in (14) is not the same partitioning used in (11).
Using the partitioning in (14), let us write the corresponding respective solutions to (4) and (5)
as
$$Q=\bmatrix Q_1&Q_2\\
\noalign{\medskip}
Q_3&Q_4\endbmatrix,\quad N=\bmatrix N_1&N_2\\
\noalign{\medskip}
N_3&N_4\endbmatrix,\tag 15$$
where $Q_1$ and $N_1$ have sizes $q\times q,$ $Q_4$ and $N_4$ have sizes $(p-q)\times (p-q),$
etc.
Note that because of the selfadjointness of $Q$ and $N$ stated in Theorem 1, we have
$$Q_1^\dagger=Q_1,\quad
Q_2^\dagger=Q_3,\quad Q_4^\dagger=Q_4,\quad
N_1^\dagger=N_1,\quad
N_2^\dagger=N_3,\quad N_4^\dagger=N_4.$$
Furthermore, from Theorem 3 it follows that $Q_1,$ $Q_4$, $N_1,$ and
$N_4$ are all invertible.

\noindent{\bf Theorem 6.} {\it Assume that the triplet $(A,B,C)$ partitioned as in (14)
corresponds to a minimal realization in (3) and that all eigenvalues of $A$
have positive real parts. Consider the transformation (12)
with $(\tilde A,\tilde B,\tilde C)$ having similar
block representations as in (14), $(Q,N)$ as in (15)
corresponding to the unique solution to
the Lyapunov system in (4) and (5),}
$$\tilde A_1=A_1,\quad \tilde A_2=-A_2^\dagger,\quad
\tilde B_1=B_1-N_2N_4^{-1}B_2,\quad
\tilde B_2=-N_4^{-1}B_2,$$
$$\tilde C_1=C_1-C_2Q_4^{-1}Q_3,\quad
\tilde C_2=-C_2Q_4^{-1},$$
{\it and $(\tilde Q,\tilde N)$ given as}
$$\tilde Q_1=Q_1-Q_2Q_4^{-1}Q_3,\quad
\tilde Q_2=-Q_2Q_4^{-1},\quad
\tilde Q_3=-Q_4^{-1}Q_3,\quad
\tilde Q_4=-Q_4^{-1},$$
$$\tilde N_1=N_1-N_2N_4^{-1}N_3,\quad \tilde N_2=-N_2N_4^{-1},\quad
\tilde N_3=-N_4^{-1}N_3,\quad
\tilde N_4=-N_4^{-1},$$
{\it and $\tilde F$ and $\tilde u$
as in (8) and (9), respectively,
but by using $(\tilde A,\tilde B,\tilde C,\tilde Q,\tilde N)$ instead of $(A,B,C,Q,N)$
on the right hand sides. We then have the following:}

\item{(i)} {\it The matrices $\tilde Q$ and $\tilde N$ are selfadjoint and invertible.
They satisfy the respective
Lyapunov equations in (13).}

\item{(ii)} {\it The quantity $F$ is transformed
according to}
$$\tilde F=\bmatrix I&-Q_2Q_4^{-1}\\
\noalign{\medskip}
0&-Q_4^{-1}\endbmatrix F
\bmatrix I&0\\
\noalign{\medskip}
-N_4^{-1}N_3&-N_4^{-1}\endbmatrix,$$
\item{} {\it and the matrix $\tilde F$ is invertible
at every point on the $xt$-plane.}
\item{(iii)} {\it The triplets $(A,B,C)$ and $(\tilde A,\tilde B,\tilde C)$
are equivalent; i.e.
$\tilde u(x,t)=u(x,t).$}

\vskip 10 pt
\noindent {\bf 3. EXAMPLES}
\vskip 3 pt

In this section we illustrate our method of constructing exact solutions
to the NLS equation with some concrete examples.

\noindent {\bf Example 1.} The well-known ``$n$-soliton" solution to
the NLS equation
is obtained
by choosing the triplet $(A,B,C)$ as
$$A=\text{diag}\{a_1, a_2,\dots,a_n\},\quad
B^\dagger=\bmatrix 1&1&\dots& 1\endbmatrix,\quad C=\bmatrix
c_1&c_2&\dots&c_n\endbmatrix,$$ where $a_j$ are (complex) nonzero
constants with positive real parts, $B$ contains $n$ entries,
and the quantities $c_j$ are complex constants. Note that diag
is used to denote the diagonal matrix. In this case, using (8)
and (10) we evaluate the $(j,k)$-entries of the $n\times n$
matrix-valued functions $Q,$ $N,$ and $F(x,t)$ as
$$N_{jk}=\displaystyle\frac{1}{a_j+a_k^*},\quad
Q_{jk}=\displaystyle\frac{c_j^* c_k}{a_j^*+a_k},\quad
F_{jk}=\delta_{jk}e^{2a_j^*x-4i (a_j^*)^2 t}+
\displaystyle\sum_{s=1}^n \frac{c_j^* c_s\,e^{-2a_sx-4i a_s^2
t}}{(a_j^*+a_s)(a_s+a_k^*)},
$$
where $\delta_{jk}$ denotes the Kronecker delta. Having obtained
$Q,$ $N,$ and $F(x,t),$
we construct the solution $u(x,t)$ to the NLS equation
via (9) or equivalently as the ratio of two determinants as
$$u(x,t)=\displaystyle\frac{2}{\det F(x,t)}\left|\matrix 0&B^\dagger\\
\noalign{\medskip}
C^\dagger&F(x,t)\endmatrix
\right|
.\tag 16$$
For example, when $n=1,$ from (16) we obtain the single soliton solution
$$u(x,t)=\displaystyle\frac{-8c_1^* (\text{Re}[a_1])^2 \,e^{-2a_1^*
x+4i (a_1^*)^2t}} {4
(\text{Re}[a_1])^2+|c_1|^2\,e^{-4x(\text{Re}[a_1])+8t
(\text{Im}[a_1^2])}},$$ where Re and Im denote the real and
imaginary parts, respectively. From (1) we see that if $u(x,t)$
is a solution to (1), so is $e^{i\theta}u(x,t)$ for any real
constant $\theta.$ Hence, the constant phase factor
$e^{i\theta}$ can always be omitted from the solution to (1)
without any loss of generality. As a result, we can write the
single soliton solution also in the form
$$u(x,t)=2\,\text{Re}[a_1]\,
\,e^{-2ix \text{Re}[a_1]+4it\,\text{Re}[a_1^2]}
\text{sech}\left(2\,\text{Re}[a_1](x-4 t\,\text{Im}[a_1])
-\log|c_1/2 \text{Re}[a_1]|\right) ,$$ where it is seen that
$u(x,t)$ has amplitude $2\,\text{Re}[a_1]$ and moves with
velocity $4\,\text{Im}[a_1].$

\noindent {\bf Example 2.}
For the triplet $(A,B,C)$ given by
$$A=\bmatrix 2&0\\
\noalign{\medskip}
0&-1\endbmatrix,\quad
B=\bmatrix 1\\
\noalign{\medskip}
1\endbmatrix,\quad
C=\bmatrix 1&-1\endbmatrix,\tag 17$$
we evaluate $Q$ and $N$ explicitly by solving (4) and (5),
respectively, as
$$
N=\bmatrix 1/4&1\\
\noalign{\medskip}
1&-1/2\endbmatrix,\quad
Q=\bmatrix 1/4&-1\\
\noalign{\medskip}
-1&-1/2\endbmatrix,$$
and obtain $F(x,t)$ by using (8) as
$$F(x,t)=\bmatrix
e^{4x-16it}-e^{2x-4it}+\displaystyle\frac{1}{16}\,e^{-4x-16it}&
\displaystyle\frac{1}{4}\,e^{-4x-16it}
+\displaystyle\frac{1}{2}\,e^{2x-4it}\\
\noalign{\medskip}
-\displaystyle\frac{1}{4}\,e^{-4x-16it}
-\displaystyle\frac{1}{2}\,e^{2x-4it}&
e^{-2x-4it}-e^{-4x-16it}+\displaystyle\frac{1}{4}\,e^{2x-4it}\endbmatrix$$
Finally, using (9), we obtain the corresponding
solution to the NLS equation as
$$u(x,t)=\displaystyle\frac{8 e^{4 i t}(9e^{-4 x}+16 e^{4 x})-
32e^{16i t}(4e^{-2 x}+9 e^{2 x})} {-128\cos (12 t)+4e^{-6 x}+16e^{6
x}+81e^{-2 x}+64e^{2 x}}.\tag 18$$
It can independently be verified that $u(x,t)$ given
in (18) satisfies the NLS equation on the entire $xt$-plane.

With the help of the results stated in Section 2,
we can determine triplets $(\tilde A,\tilde B,\tilde C)$
that are equivalent to the triplet in (17). The following triplets
all yield the same $u(x,t)$ given in (18):

\item{(i)} $\tilde A=\bmatrix 2&0\\
0&1\endbmatrix,\
\tilde B=\bmatrix 9/\alpha_1\\
-4/\alpha_2\endbmatrix,\
\tilde C=\bmatrix \alpha_1&\alpha_2\endbmatrix,$ where $\alpha_1$ and $\alpha_2$
are arbitrary (complex)
nonzero
parameters. Note that both eigenvalues of $\tilde A$ are positive, whereas
only one of the eigenvalues of $A$ in (17) is positive.

\item{(ii)} $\tilde A=\bmatrix -2&0\\
0&1\endbmatrix,\
\tilde B=\bmatrix 16/(9\alpha_3)\\
-4/(9\alpha_4)\endbmatrix,\
\tilde C=\bmatrix \alpha_3&\alpha_4\endbmatrix,$
where $\alpha_3$ and $\alpha_4$ are arbitrary (complex)
nonzero
parameters. Note that the eigenvalues of $\tilde A$ in this
triplet are negatives of the eigenvalues of $A$ given in (17).

\item{(iii)} $\tilde A=\bmatrix 2&0\\
0&-1\endbmatrix,\
\tilde B=\bmatrix 1/\alpha_5\\
-1/\alpha_6\endbmatrix,\ \tilde C=\bmatrix
\alpha_5&\alpha_6\endbmatrix,$ where $\alpha_5$ and $\alpha_6$
are arbitrary (complex) nonzero parameters. Note that $\tilde
A$ here agrees with $A$ in (17).

\item{(iv)} $\tilde A=\bmatrix -2&0\\
0&-1\endbmatrix,\
\tilde B=\bmatrix 16/\alpha_7\\
-9/\alpha_8\endbmatrix,\
\tilde C=\bmatrix \alpha_7&\alpha_8\endbmatrix,$ where $\alpha_7$ and $\alpha_8$ are
arbitrary (complex)
nonzero
parameters. Note that both eigenvalues of $\tilde A$ are negative.

\item{(v)} Equivalent to (17) we also have the triplet $(\tilde A,\tilde B,\tilde C)$
given by
$$\tilde A=\bmatrix \alpha_9&\alpha_{10}\\
\noalign{\medskip}
\displaystyle\frac{(1-\alpha_9)(\alpha_9-2)}{\alpha_{10}}
&3-\alpha_9\endbmatrix,\quad
\tilde C=\bmatrix \alpha_{11}&\alpha_{12}\endbmatrix,$$
$$
\tilde B=\displaystyle\frac{\bmatrix 5\alpha_{10}^2\alpha_{11}+\alpha_{10}\alpha_{12}
-5\alpha_9\alpha_{10}\alpha_{12}\\
\noalign{\medskip}
14\alpha_{10}\alpha_{11}-5\alpha_9\alpha_{10}\alpha_{11}
+10\alpha_{12}-15\alpha_9\alpha_{12}+5\alpha_9^2
\alpha_{12}\endbmatrix}{\alpha_{10}^2\alpha_{11}^2+3\alpha_{10}
\alpha_{11}\alpha_{12}-2\alpha_9\alpha_{10}\alpha_{11}\alpha_{12}+
2\alpha_{12}^2-3\alpha_9\alpha_{12}^2+\alpha_9^2\alpha_{12}^2}
,$$ where $\alpha_9,\dots,\alpha_{12}$ are arbitrary parameters
with the restriction that $\alpha_{10}\alpha_{11}\alpha_{12}\ne
0,$ which guarantees that the denominator of $\tilde B$ is
nonzero; when $\alpha_{10}=0$ we must have
$\alpha_{11}\alpha_{12}\ne 0$ and choose $\alpha_9$ as $2$ or
$1.$ In fact, the minimality of the triplet $(\tilde A,\tilde
B,\tilde C)$ guarantees that $\tilde B$ is well defined. For
example, the triplet is not minimal if
$\alpha_{11}\alpha_{12}=0.$ We note that the eigenvalues of
$\tilde A$ are $2$ and $1$ and that $\tilde A$ here is similar
to the matrix $\tilde A$ in the equivalent triplet given in
(i).

Other triplets equivalent to (17) can be found as in (v) above, by exploiting the
similarity for the matrix $\tilde A$ given in (ii), (iii), and (iv), respectively, and
by using (3) to determine the corresponding $\tilde B$ and $\tilde C$ in the triplet.

\vskip 10 pt

\noindent{\bf Acknowledgments.}
The research leading to this article
was supported in part by the U.S. National Science Foundation under grant
DMS-0610494, the
Italian Ministry of Education and Research (MIUR) under PRIN grant no.
2006017542-003, and INdAM-GNCS.

\vskip 5 pt

\noindent {\bf{REFERENCES}}

\item{[1]} M. J. Ablowitz and P. A. Clarkson, {\it Solitons, nonlinear
evolution equations and inverse scattering,} Cambridge Univ. Press, Cambridge,
1991.

\item{[2]} M. J. Ablowitz and H. Segur, {\it
Solitons and the inverse scattering
transform,} SIAM, Philadelphia, 1981.

\item{[3]} T. Aktosun, T. Busse, F. Demontis, and C. van der Mee,
{\it Symmetries for exact solutions to the nonlinear Schr\"odinger equation,} preprint,
arXiv: 0905.4231.

\item{[4]} T. Aktosun, F. Demontis, and C. van der Mee,
{\it Exact solutions to the focusing nonlinear Schr\"odinger equation,}
Inverse Problems {\bf 23}, 2171--2195 (2007).

\item{[5]} T. Aktosun and C. van der Mee, {\it Explicit solutions to the
Korteweg-de Vries equation on the half-line,} Inverse Problems {\bf 22},
2165--2174 (2006).

\item{[6]} H. Bart, I. Gohberg, M. A. Kaashoek, and A. C. M. Ran,
{\it Factorization of matrix and operator functions.
The state space method,} Birkh\"auser,
Basel, 2007.

\item{[7]} T. Busse, Ph.D. thesis, University of Texas at Arlington,
2008.

\item{[8]} F. Demontis, {\it Direct and inverse scattering of the matrix
Zakharov-Shabat system}, Ph.D. thesis, University of Cagliari, Italy, 2007.

\item{[9]} A. Hasegawa and M. Matsumoto, {\it Optical solitons in fibers,}
3rd ed., Springer, Berlin, 2002.

\item{[10]} A. Hasegawa and F. Tappert, {\it Transmission of stationary
nonlinear optical pulses in dispersive dielectric fibers. I. Anomalous
dispersion,} Appl. Phys. Lett. {\bf 23}, 142--144 (1973).

\item{[11]} A. Hasegawa and F. Tappert, {\it Transmission of stationary
nonlinear optical pulses in dispersive dielectric fibers. II. Normal
dispersion,} Appl. Phys. Lett. {\bf 23}, 171--172 (1973).

\item{[12]} S. Novikov, S. V. Manakov, L. P. Pitaevskii, and V. E. Zakharov,
{\it Theory of solitons,}
Consultants Bureau, New York, 1984.

\item{[13]} E. Olmedilla, {\it Multiple pole solutions of the nonlinear
Schr\"odinger equation,} Phys. D {\bf 25}, 330--346 (1987).

\item{[14]} V. E. Zakharov, {\it Stability of periodic waves of
finite amplitude on the surface of a deep fluid,}
J. Appl. Mech. Tech. Phys. {\bf 4}, 190--194 (1968).

\item{[15]} V. E. Zakharov and A. B. Shabat, {\it Exact theory of
two-dimensional self-focusing and one-dimensional self-modulation of waves in
nonlinear media,} Sov. Phys. JETP {\bf 34}, 62--69 (1972).

\end

he triplet $(A,B,C)$ in Example 2 corresponds to a minimal
We note that the solution in Example 2 is not
a soliton solution to the NLS equation because one of the eigenvalues
of the matrix $A$ is negative. Its Mathematica animation is available [15],
and in Figure 1 we provide some snapshots of
$|u(x,t)|.$

\vskip 20 pt

\centerline{\hbox{\psfig{figure=ex7.2fort=0.0.ps,width=2.2 truein,height=1.3 truein}}
\ {\psfig{figure=ex7.2fort=0.1.ps,height=1.3 truein,width=2.2 truein}}
\ {\psfig{figure=ex7.2fort=0.2.ps,height=1.3 truein,width=2.2 truein}}}

\centerline{\hbox{\psfig{figure=ex7.2fort=0.3.ps,width=2.2 truein,height=1.3 truein}}
\ {\psfig{figure=ex7.2fort=0.4.ps,height=1.3 truein,width=2.2 truein}}
\ {\psfig{figure=ex7.2fort=0.5.ps,height=1.3 truein,width=2.2 truein}}}

\centerline{{\bf Figure~1.} Snapshots of $|u(x,t)|$
of Example 2 at $t=0.0,$ $0.1,$
$0.2,$ $0.3,$ $0.4,$ and $0.5.$}

\vskip 10 pt

\noindent {\bf Example 3.}
Letting
$$A=\bmatrix 2-i&-1\\
0&2-i\endbmatrix,\quad
B=\bmatrix 0\\
1\endbmatrix,\quad
C=\bmatrix 1+2i&-1+4i\endbmatrix,$$
we evaluate
$P(x;t)$ via (4),
$N$ via (7) or (8), and $Q$ via (6) or (8). We get
$$P(x;t)=e^{-(4-2i)x-(16+12i)t} \bmatrix 1
&2x+(8+16i)t\\
0&1\endbmatrix,$$
$$
N=\bmatrix 1/32&1/16\\
1/16&1/4\endbmatrix,\quad
Q=\bmatrix 5/4&33/16+3i/2\\
33/16-3i/2&169/32\endbmatrix.$$
Using (14)-(16) in (10), we obtain the corresponding
solution to the NLS equation as
$u(x,t)=\text{num}(x,t)/\text{den}(x,t),$ where
$$\aligned
\text{num}(x,t):=&
1024 e^{4(x+4 t)-2i(x-6t)}\left[
(12-9i)+100t+(5-10i)x\right]\\
&+131072e^{12(x+4 t)-2i(x-6t)}\left[
(1+4i)+(24+32i)t-(2-4i)x\right],\endaligned$$
$$\aligned\text{den}(x,t):=&25+65536
   e^{16 (4 t+x)}
   \\
   &+512 e^{8 (4 t+x)} \left[
   12800
   t^2+64 (20 x+43) t+160 x^2+304 x+207\right].
\endaligned$$
The solution in Example 3 can be described as a soliton of double
multiplicity, and its Mathematica animation is available [15].
Some snapshots of $|u(x,t)|$ are shown in Figure~2.

\vskip 20 pt

\centerline{\hbox{\psfig{figure=ex7.3fort=-0.5.ps,width=2.2 truein,height=1.3 truein}}
\ {\psfig{figure=ex7.3fort=-0.2.ps,height=1.3 truein,width=2.2 truein}}
\ {\psfig{figure=ex7.3fort=-0.1.ps,height=1.3 truein,width=2.2 truein}}}

\centerline{\hbox{\psfig{figure=ex7.3fort=0.0.ps,width=2.2 truein,height=1.3 truein}}
\ {\psfig{figure=ex7.3fort=0.1.ps,height=1.3 truein,width=2.2 truein}}
\ {\psfig{figure=ex7.3fort=0.2.ps,height=1.3 truein,width=2.2 truein}}}

\centerline{{\bf Figure 2.} Snapshots of $|u(x,t)|$
of Example 3 at $t=-0.5,$ $-0.2,$
$-0.1,$ $0.0,$ $0.1,$ and $0.2.$}

\vskip 10 pt

\noindent {\bf Example 4.}
As our input let us use
$$A=\bmatrix 1&-1&0\\
0&1&-1\\
0&0&1\endbmatrix,\quad
B=\bmatrix 0\\
0\\
1\endbmatrix,\quad
C=\bmatrix 1&0&0\endbmatrix,\tag 17$$
and evaluate
$P(x;t)$ via (4),
$N$ via (7) or (8), and $Q$ via (6) or (8). We get
$$P(x;t)=e^{-2x-4it} \bmatrix 1
&2x+8it&2x^2-32t^2+4it(4x-1)\\
0&1&2x+8it\\
0&0&1\endbmatrix,\tag 18$$
$$
N=\bmatrix 3/16&3/16&1/8\\
3/16&1/4&1/4\\
1/8&1/4&1/2\endbmatrix,\quad
Q=\bmatrix 1/2&1/4&1/8\\
1/4&1/4&3/16\\
1/8&3/16&3/16\endbmatrix.\tag 19$$
The use of (17)-(19) in (10) results in a
solution to the NLS equation which is given by
$u(x,t)=\text{num}(x,t)/\text{den}(x,t),$ where
$$\aligned
\text{num}(x,t)&:=32e^{-2(x-2it)}\left\{[-32768x^2+524288t^2+262144itx-65536it]
\right.\\ &+\left.e^{-4x}\left[90112t^2+15872x^2+131072t^2x^2+4096x^4
+196608xt^2\right.\right.\\ &+\left.\left.12288x^3+9216x+1344+1048576t^4
-32768itx^2-35840it\right.\right.\\ &-\left.\left.61440itx\right]
+e^{-8x}\left[128t^2-8x^2-24x-15-112it-64itx\right]\right\},\endaligned$$
$$\aligned\text{den}(x,t)&:=262144+e^{-4x}\left[262144x^4+589824x^2+393216x+524288x^3
\right.\\ &+\left.67108864t^4+8388608x^2t^2+122880\right]
+e^{-8x}\left[16384x^3+4096x^4\right.\\ &+\left.1048576t^4+15360x+344064t^2
+24576x^2+131072x^2t^2\right.\\ &+\left.393216xt^2+3648\right]+e^{-12x}.
\endaligned$$
The solution in Example 4 can be described as a single soliton of triple
multiplicity. A Mathematica notebook [15] is available for this
example with a corresponding animation
of $|u(x,t)|.$ In Figure 3 we present some snapshots of $|u(x,t)|.$

\vskip 20 pt

\centerline{\hbox{\psfig{figure=ex7.4fort=0.0.ps,width=2.2 truein,height=1.3 truein}}
\ {\psfig{figure=ex7.4fort=0.1.ps,height=1.3 truein,width=2.2 truein}}
\ {\psfig{figure=ex7.4fort=0.2.ps,height=1.3 truein,width=2.2 truein}}}

\centerline{\hbox{\psfig{figure=ex7.4fort=0.3.ps,width=2.2 truein,height=1.3 truein}}
\ {\psfig{figure=ex7.4fort=0.4.ps,height=1.3 truein,width=2.2 truein}}
\ {\psfig{figure=ex7.4fort=0.5.ps,height=1.3 truein,width=2.2 truein}}}

\centerline{{\bf Figure 3.} Snapshots of $|u(x,t)|$
of Example 4 at $t=0.0,$ $0.1,$
$0.2,$ $0.3,$ $0.4,$ and $0.5.$}

\vskip 10 pt

\noindent{\bf Acknowledgments.}
The research leading to this article
was supported in part by the U.S. National Science Foundation under grant
DMS-0610494, the
Italian Ministry of Education and Research (MIUR) under PRIN grant no.
2006017542-003, and INdAM-GNCS.

\vskip 5 pt

\noindent {\bf{REFERENCES}}

\item{[8]} M. J. Ablowitz and P. A. Clarkson, {\it Solitons, nonlinear
evolution equations and inverse scattering,} Cambridge Univ. Press, Cambridge,
1991.

\item{[9]} M. J. Ablowitz and H. Segur, {\it
Solitons and the inverse scattering
transform,} SIAM, Philadelphia, 1981.

\item{[1a]} T. Aktosun, T. Busse, F. Demontis, and C. van der Mee,
{\it Symmetries for exact solutions to the nonlinear Schr\"odinger equation,} preprint,
arXiv: 0905.4231.

\item{[1]} T. Aktosun, F. Demontis, and C. van der Mee,
{\it Exact solutions to the focusing nonlinear Schr\"odinger equation,}
Inverse Problems {\bf 23}, 2171--2195 (2007).

\item{[11]} T. Aktosun and C. van der Mee, {\it Explicit solutions to the
Korteweg-de Vries equation on the half-line,} Inverse Problems {\bf 22},
2165--2174 (2006).

\item{[7]} H. Bart, I. Gohberg, M. A. Kaashoek, and A. C. M. Ran,
{\it Factorization of matrix and operator functions.
The state space method,} Birkh\"auser,
Basel, 2007.

\item{[14]} T. Busse, Ph.D. thesis, University of Texas at Arlington,
2008.

\item{[12]} F. Demontis, {\it Direct and inverse scattering of the matrix
Zakharov-Shabat system}, Ph.D. thesis, University of Cagliari, Italy, 2007.

\item{[4]} A. Hasegawa and M. Matsumoto, {\it Optical solitons in fibers,}
3rd ed., Springer, Berlin, 2002.

\item{[5]} A. Hasegawa and F. Tappert, {\it Transmission of stationary
nonlinear optical pulses in dispersive dielectric fibers. I. Anomalous
dispersion,} Appl. Phys. Lett. {\bf 23}, 142--144 (1973).

\item{[6]} A. Hasegawa and F. Tappert, {\it Transmission of stationary
nonlinear optical pulses in dispersive dielectric fibers. II. Normal
dispersion,} Appl. Phys. Lett. {\bf 23}, 171--172 (1973).

\item{[10]} S. Novikov, S. V. Manakov, L. P. Pitaevskii, and V. E. Zakharov,
{\it Theory of solitons,}
Consultants Bureau, New York, 1984.

\item{[13]} E. Olmedilla, {\it Multiple pole solutions of the nonlinear
Schr\"odinger equation,} Phys. D {\bf 25}, 330--346 (1987).

\item{[3]} V. E. Zakharov, {\it Stability of periodic waves of
finite amplitude on the surface of a deep fluid,}
J. Appl. Mech. Tech. Phys. {\bf 4}, 190--194 (1968).

\item{[2]} V. E. Zakharov and A. B. Shabat, {\it Exact theory of
two-dimensional self-focusing and one-dimensional self-modulation of waves in
nonlinear media,} Sov. Phys. JETP {\bf 34}, 62--69 (1972).

\end